\documentclass[prb,twocolumn,showpacs,superscriptaddress]{revtex4}

\usepackage{graphicx}
\usepackage{amsmath}
\usepackage{amssymb}
\usepackage{epsfig}

\renewcommand{\v}[1]{{\bf #1}}

\def\eqa{\begin{eqnarray}}
\def\eea{\end{eqnarray}}
\newcommand{\eq}{\begin{equation}}
\newcommand{\ee}{\end{equation}}
\newcommand{\nn}{\nonumber\\}
\newcommand{\Eq}[1]{Eq.~(\ref{#1})}
\newcommand{\<}{\langle}
\renewcommand{\>}{\rangle}

\newcommand{\p}{\partial}

\newcommand{\ua}{\uparrow}
\newcommand{\da}{\downarrow}
\newcommand{\ra}{\rightarrow}

\newcommand{\eps}{\epsilon}

\newcommand{\Ga}{\Gamma}

\newcommand{\La}{\Lambda}

\newcommand{\si}{\sigma}

\usepackage{color}
\begin{document}
\title{Electronic orders near the type-II van Hove singularity in BC$_3$}

\author{Yao Wang}
\affiliation{National Laboratory of Solid State Microstructures $\&$ School of Physics, Nanjing
University, Nanjing, 210093, China}

\author{Jin-Guo Liu}
\affiliation{Beijing National Laboratory for Condensed Matter Physics and Institute of Physics, Chinese Academy of Sciences, Beijing 100190, China}

\author{Wan-Sheng Wang}
\affiliation{Department of Physics, Ningbo University, Ningbo 315211, China}

\author{Qiang-Hua Wang}
\affiliation{National Laboratory of Solid State Microstructures $\&$ School of Physics, Nanjing
University, Nanjing, 210093, China}
\affiliation{Collaborative Innovation Center of Advanced Microstructures, Nanjing 210093, China}
\email{qhwang@nju.edu.cn}
%\date{\today}

\begin{abstract}

Using the functional renormalization group, we investigate the electron instability in the single-sheet BC$_3$ when the electron filling is near a type-II van Hove singularity. For a finite Hubbard interaction, the ferromagnetic-like spin density wave order dominates in the immediate vicinity of the singularity. Elsewhere near the singularity the $p$-wave superconductivity prevails. We also find that a small nearest-neighbor Coulomb repulsion can enhance the superconductivity. Our results show that BC$_3$ would be a promising candidate to realize topological $p+ip'$ superconductivity, but the transition temperature is practically sizable only if the local interaction is moderately strong.

\end{abstract}

%
%75.30.Fv  Spin-density waves
%74.20.Rp  Pairing symmetries (other than s-wave)
%74.20.-z  Theories and models of superconducting state
%71.27.+a  Strongly correlated electron systems; heavy fermions
%64.60.ae  Renormalization-group theory

\maketitle

\section{Introduction}

The Majorana modes in the vortices of topological superconductors can be used in quantum computation,
motivating continuing efforts in searching for topological superconductors. \cite{Sato2016} The leading candidate for intrinsic topological superconductors is Sr$_2$RuO$_4$, which may host triplet $p+ip'$-wave superconductivity.\cite{Maeno2012} In this material, the electron filling is near the van Hove singularity. This singularity is classified as of type-I, in the sense that the van Hove momenta are time-reversal invariant. Because of the proximity to the van Hove singularity, there are strong ferromagnetic fluctuations at low energy. This may either lead to ferromagnetic order, or trigger triplet pairing. At a first sight, by doping up to the van Hove level, the logarithmically diverging density of states at the Fermi level would enhance ferromagnetic correlations and possibly triplet pairing also. However, a pair of opposite time-reversal invariant momenta correspond to the same lattice momentum, and hence equal-spin pairing on such momenta is forbidden by Pauli exclusion principle. This would be a destructive effect for triplet pairing. An interesting recent proposal~\cite{Yao2015} is to look for materials in which the van Hove momenta are not time-reversal invariant and hence the above destructive effect is avoided. This kind of van Hove singularity is classified as of type-II.

The single-sheet BC$_3$ is a suitable material with type-II van Hove singularity (see Fig.\ref{model}). It is a graphene-like material with a layered hexagonal structure.\cite{Tang2013} According to first-principle calculations, \cite{Miyamoto1994} the undoped BC$_3$ is a semimetal with a band gap about 0.54 eV. The first and second conduction bands are $\pi$ and $\pi^\ast$ bands, which cross at $K$ and $K'$ in the Brillouine zone but are isolated from the other bands.  A macroscopic uniform sheet of single-crystal BC$_3$ was reported to be available by carbon-substitution in a boron honeycomb. \cite{Tanaka2005} Superconductivity is theoretically considered in terms of electron-phonon coupling,\cite{Cohen2011} but no trace of superconductivity has been found under hole-doping.\cite{Ueno2006} However, by electron doping (into the $\pi$ band), strong ferromagnetic fluctuations were predicted and may be attributed to the proximity to the van Hove singularity. \cite{Chen2013}  This makes electron-doped BC$_3$ a hopeful candidate for triplet pairing.

Previously, the possible superconductivity was analyzed by renormalization group (RG) in the limit of weak coupling right at the van Hove level, and random-phase approximation based fluctuation-exchange (FLEX) approximation slightly away from the van Hove level. \cite{Chen2015} Both schemes predict that $p$-wave pairing is the leading instability for weak interactions. Here we ask how it fairs as the interaction becomes stronger so that competing or interwinning orders have to be addressed. We apply the singular-mode functional renormalization group (SM-FRG) \cite{Wang2012,Xiang2012,Yang2013,Wang2013,Wang2016,Liu2017} that can treat all channels on equal footing.

Our results at weak coupling is consistent with RG and FLEX in Ref.\cite{Chen2015}. However, for moderate Hubbard interactions, the ferromagnetic or ferromagnetic-like spin-density wave order dominates in the immediate vicinity of the van Hove singularity, while $p$-wave pairing is present elsewhere near the singularity. The transition temperature becomes practically sizable (of the order of Kelvin) only when the local Hubbard interaction is several times stronger than that estimated by first principle calculations. We also find a weak nearest-neighbor Coulomb repulsion can enhance the transition temperature significantly. Our results call for refined estimation of the interaction parameters before one can decide whether BC$_3$ would be of practical interest as a $p$-wave superconductor.

The rest of the paper is as follows. We introduce the model and the SM-FRG in Sec.\ref{M&M}. The results are presented and discussed in Sec.\ref{R&D}. Finally, Sec.\ref{SMR} is a summary of this work.

\section{Model and Method} \label{M&M}

\begin{figure}
\includegraphics[width=8.5cm,trim={7.5cm 0.5cm 7.5cm 1cm},clip]{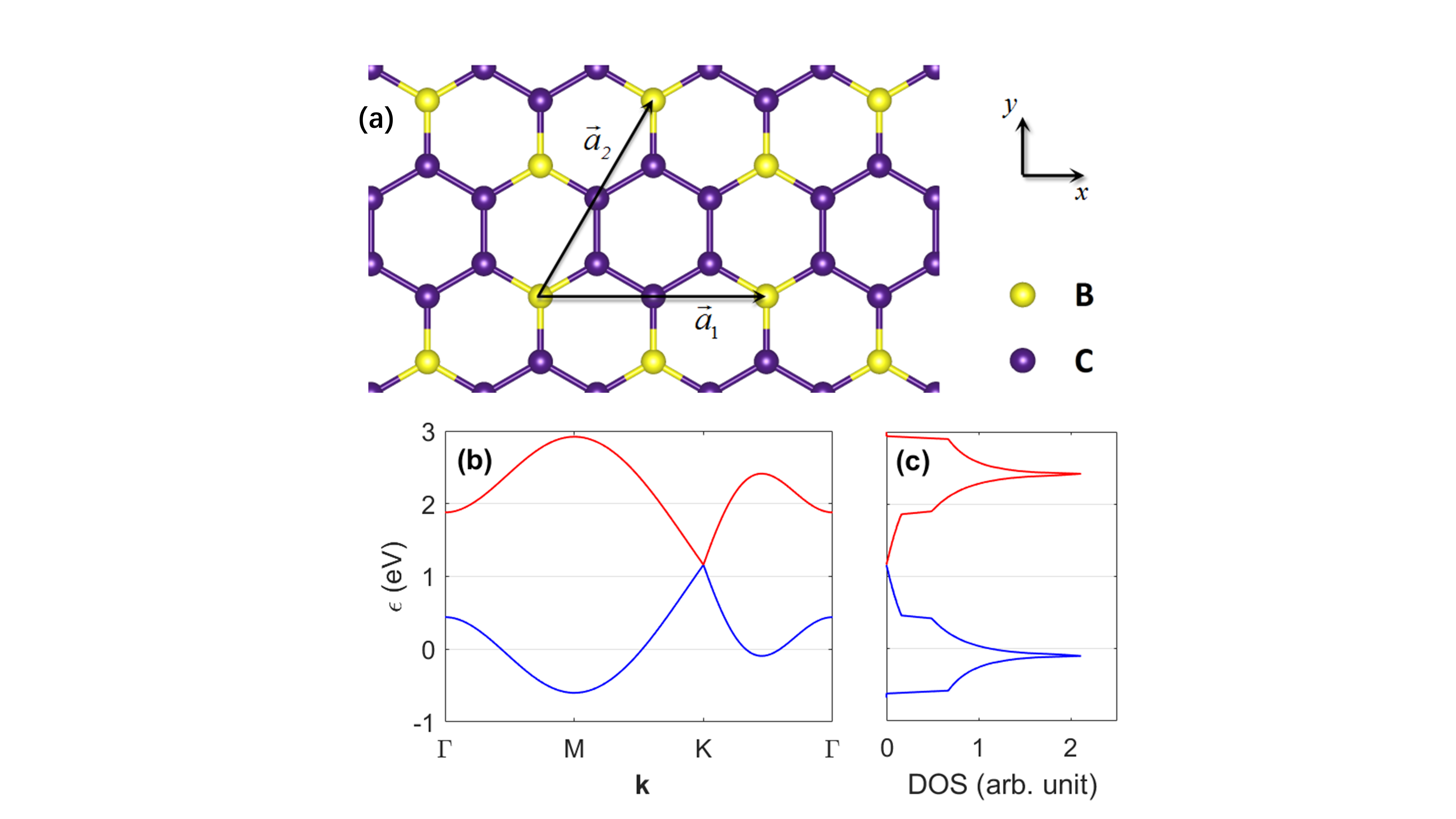}
\caption{ (Color online) (a) Structure of BC$_3$ monolayer. Here $\v a_1$ and $\v a_2$ are basis vectors. Purple (yellow) ball represents carbon (boron) atom. (b) Band energy $\eps$ as a function of momentum $\v k$ along high symmetry cuts in the Brillouine zone. (c) Density of states.}\label{model}
\end{figure}

The structure of monolayer BC$_3$ is shown in Fig.\ref{model}(a). Every boron atom is connected to three carbon atoms, and every boron hexagon encloses a smaller carbon hexagon. The conduction bands are mainly derived from the $p_z$ orbital of boron, justifying a simplified model for borons on a honeycomb lattice, with the following Hamiltonian,
\eqa
H = &&-\sum_{\< ij \> \si}(c^\dag_{i\si}t_{ij}c_{j\si}+{\rm h.c.})-\mu\sum_{i\si}n_{i\si}\nn
&&+U\sum_i n_{i\ua}n_{i\da}+V\sum_{\< ij \>\in {\rm NN}}n_i n_j. \label{H}
\eea
where $c^\dag_{i\sigma}$ is the annihilation operator at site $i$ with spin $\sigma$, $\<ij\>$ denotes first-, second- and third-neighbor bonds, and $\mu$ is the chemical potential. According to a fit to the first-principle band structure,  $t_{1} = 0.62 $ eV, $t_{2} = 0$ and $t_{3}=-0.38 $ eV for the first, second and third neighbors. The onsite Hubbard interaction was estimated as $U\sim 0.7 $ eV, \cite{Chen2015} but for systematics we leave both $U$ and the nearest-neighbor (NN) repulsion $V$ as parameters. Finally, spin-orbital coupling (SOC) may arise from the missing mirror symmetry about the second-neighbor boron-boron bonds, but this SOC is expected to be weak given the light elements and the relatively long bonds. For this reason, we will ignore SOC henceforth. The band dispersion along high symmetry cuts is plot in Fig.\ref{model}(b), together with the density of states (DOS) in (c). It is seen that the DOS diverges logarithmically at the van Hove energy level. (The singularity is cutoffed by the smearing factor used in the numerical calculation of DOS.) Proximity to such a singularity makes the system susceptible to various instabilities under the electron-electron interactions.

In order to treat all possible and competing electronic orders on equal footing, we apply the singular-mode functional renormalization group (SM-FRG).  Here we outline the necessary ingredients and notations, leaving technical details in the Appendix. In a nutshell, the idea is to get momentum-resolved running pseudo-potential $\Ga_{1234}$, as in $(1/2)c_{1\si}^\dagger c_{2\si'}^\dagger \Ga_{1234} c_{3\si'} c_{4\si}$, to act on low-energy fermionic degrees of freedom up to a cutoff energy scale $\La$ (for Matsubara frequency in our case). Henceforth the numerical index labels momentum/position/sublattice (but will be suppressed wherever applicable for brevity). Momentum conservation/translation symmetry is also left implicit. Starting from $\Ga$ at $\La\ra \infty$ (specified by the bare interactions $U$ and $V$), FRG generates all one-particle-irreducible corrections to $\Ga$ to arbitrary orders in the bare interactions as $\La$ decreases. Notice that $\Ga$ may evolve to be nonlocal and even diverging. To see the instability (diverging) channel, we extract at $\La$ concurrently the effective interactions in the general charge-density wave (CDW), spin-density wave (SDW) and superconductivity (SC) channels,
\eqa
&& V^{\rm CDW}_{(14)(32)} = 2 \Ga_{1234} - \Ga_{1243},\nn
&& V^{\rm SDW}_{(13)(42)} = - V^{\rm SC}_{12)(43)} = - \Ga_{1234}. \label{eq:VX}
\eea
The left-hand sides are understood as matrices with composite indices, describing scattering of fermion bilinears. Since they all originate from $\Ga$, $V^{\rm CDW/SDW/SC}$ have overlaps but are naturally treated on equal footing. We remark that the FRG flow would be equivalent to ladder or random-phase approximations in the respective channels if the overlaps were ignored in the FRG flow equation. The divergence of the leading attractive (i.e., negative) eigenvalue $S$ of $V^{\rm CDW/SDW/SC}$ decides the instability channel, the associated eigenfunction (which is a matrix in the sublattice basis) and collective momentum describe the order parameter, and the divergence energy scale $\La_c$ is representative of the transition temperature $T_c$. More technical details can be found in Refs.\cite{Wang2012, Xiang2012, Yang2013, Wang2013, Wang2016, Liu2017} and also in the Appendix for self completeness.

\section{Results and Discussions} \label{R&D}

\begin{figure}
\includegraphics[width=8.5cm,trim={0.0cm 0.0cm 0.0cm 0.0cm},clip]{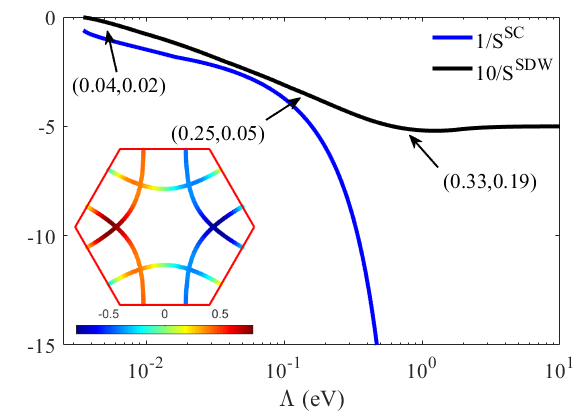}
\caption{(Color online) Flow of (the inverse of) leading eigenvalues in the SC (blue) and SDW (black) channels for $n=0.49$ and $U=2$ eV.}\label{n49}
\end{figure}

First, we consider the van Hove filling level $n = 0.49$. This case is of special interest because of the type-II singularity. In the weak coupling limit, it has been shown that $p$-wave SC wins over ferromagnetic SDW.\cite{Chen2015} We ask how it fairs in the case of finite interaction. Fig.\ref{n49} shows FRG flow of the leading eigenvalues in the SC (blue) and SDW (black) channels for $U=2$ eV. (The CDW channel is much weaker than SC and SDW channels and will be ignored henceforth.) As the energy scale $\La$ decreases, the SDW channel always dominates over the SC channel. The arrows snapshots the collective momentum $\v Q$ (divided by $\pi$) associated with the leading SDW eigenvalue. At low energy scales $\v Q\sim 0$, and the SDW channel diverges at $\La_c\sim 4\times 10^{-3}$ eV. We checked the leading eigenfunction of $V^{\rm SDW}$ to find that it corresponds to site-local spin density. This means the system enters the ferromagnetic SDW state below a transition temperature $T_c\sim \La_c$. The SC channel is triggered attractive as the SDW channel is enhanced in the intermediate energy window, and grows thereafter. The inset shows the pairing gap function (the leading eigenfunction of $V^{\rm SC}$) on the Fermi surface. The crossing points on the Fermi surface are the van Hove momenta. They are not time-reversal invariant, and hence of type-II according to Ref.\cite{Yao2015,Chen2015}. The gap function clearly has $p$-wave symmetry, and it does not vanish at the type-II van Hove momenta. By the point-group symmetry and also in our numerics, there is another degenerate $p$-wave function (not shown). Therefore, the FRG flow reveals a well-known fact that $p$-wave triplet pairing can be triggered by ferromagnetic and ferromagnetic-like spin fluctuations, and implies that if the SDW channel remains strong but does not diverge, $p$-wave SC may become the leading instability, as we show below.

\begin{figure}
\includegraphics[width=8.5cm,trim={0cm 0cm 0.0cm 0.0cm},clip]{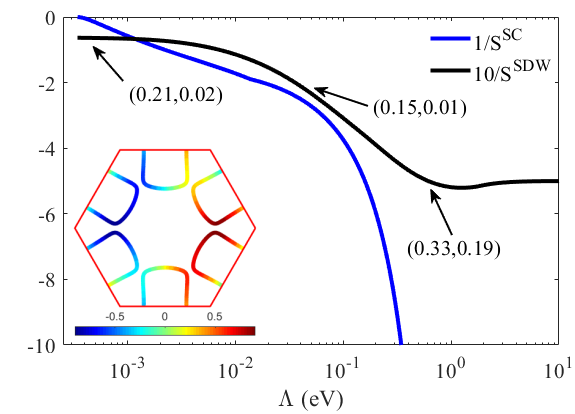}
\caption{(Color online) Flow of (the inverse of) leading eigenvalues in the SC (blue) and SDW (black) channels for $n = 0.46$ and $U=2$ eV. The inset shows one of the degenerate $p$-wave gap functions (color scale) on the Fermi surface.} \label{n46}
\end{figure}

Next we consider a filling $n = 0.46$ slightly below the van Hove level. For $U=2$ eV, the FRG flow is shown in Fig.\ref{n46}. The flow in the SDW channel is similar to that for $n=0.49$ at higher energy scales, where quasi-particle excitations are not sensitive to the fine features of the Fermi surface, but it saturates at low energy scales, with a nonzero but small $\v Q$, where SC diverges instead. The small $\v Q$ can be attributed to particle-hole scattering between the neighboring parts of the Fermi pockets. The inset shows one of the two degenerate gap functions on the Fermi surface, which is still of $p$-wave symmetry. As in the previous case, the SC channel is triggered attractive as ferromagnetic-like (or small-$\v Q$) SDW is enhanced in the intermediate energy scale. The reason that the SDW channel saturates is because the deviation from the van Hove singularity, although only slightly, regularizes the diverging density of states so that the phase space for low-energy particle-hole excitations diminishes. Instead, the SC channel, once triggered attractive, can grow with decreasing energy scale by the Cooper mechanism, even in the absence of any singularity in the density of states. This leaves room for SC to overwhelm the SDW channel. From the leading eigenfunction of $V^{\rm SC}$ we find that for the case under concern the $p$-wave pairing on NN bonds dominates, while that on longer bonds is smaller by more than one order of magnitude.  On the other hand, the two degenerate $p$-wave pairings will recombine as a fully gapped $p\pm ip'$-wave pairing in the ordered state to gain energy.

\begin{figure}
\includegraphics[width=8.5cm,trim={0cm 0cm 0.0cm 0cm},clip]{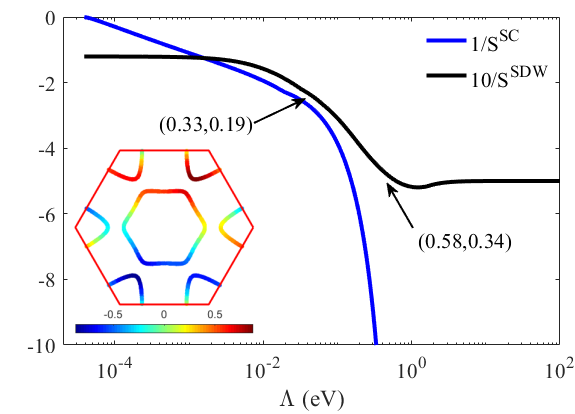}
\caption{(Color online) Flow of (the inverse of) leading eigenvalues in the SC (blue) and SDW (black) channels for $n = 0.57$ and $U=2$ eV. The inset shows one of the degenerate $p$-wave gap functions (color scale) on the Fermi surface.} \label{n57}
\end{figure}

For comparison, we also consider a filling $n=0.57$ slightly above the van Hove level. The FRG flow is shown in Fig.\ref{n57}. The overall feature is similar to that in Fig.\ref{n46}, except for the Fermi surface topology and divergence energy scale.  The difference in Fermi surface topology leads to slightly larger inter-pocket scattering momentum and hence that of the leading SDW eigenmode at the divergence scale. In Ref.\cite{Xiang2012} it is observed that proximity to van Hove singularities as well as small-momentum inter-pocket scattering are favorable to trigger triplet pairing. Interestingly both of these mechanisms are realized in proximity to a type-II van Hove singularity under concern.

\begin{figure}
    \includegraphics[width=8.5cm,trim={0cm 0cm 0cm 0cm},clip]{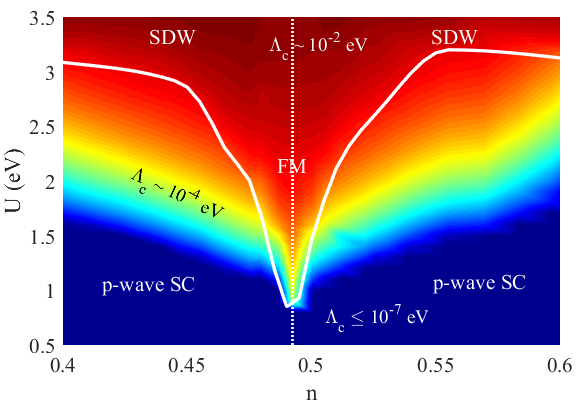}
    \caption{(Color online) Phase diagram in the $(n, U)$ parameter space. The white solid line is the phase boundary between SC and SDW phases. The dashed line indicates ferromagnetic SDW right at the van Hove filling. The color scale indicates $\log_{10}\La_c$.} \label{phasediag}
\end{figure}

\begin{figure}
    \includegraphics[width=8.5cm,trim={0cm 0cm 0cm 0cm},clip]{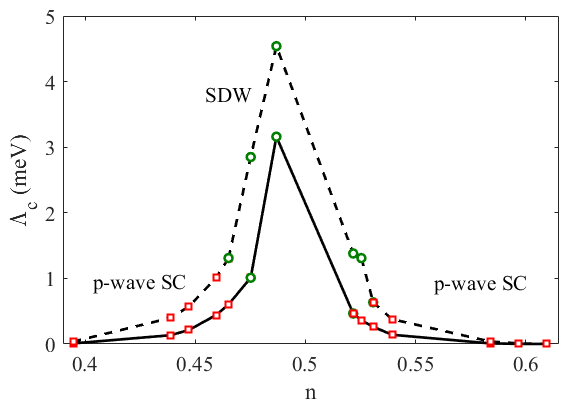}
    \caption{(Color online) Divergence scale (or transition temperature) $\La_c$ versus $n$ for $V=0$ eV (solid line) and $V=0.2$ eV (dashed line), both with $U=2$ eV. Open squares (circles) indicate $p$-wave SC (SDW) order that would emerge below $\La_c$. \label{Tc}}
\end{figure}

By systematic calculations for various values of $U$ and $n$, we obtain a phase diagram in the $(U, n)$ parameter space, as shown in Fig.\ref{phasediag}. We see that for weak interaction $U<1$ eV, the system is in the $p$-wave SC state, in agreement with the weak coupling analysis in Ref.\cite{Chen2015}, but the divergence scale (or transition temperature) is very low. For larger interactions, SDW order is favorable in the immediate vicinity of the van Hove level, with strict ferromagnetic order right at the van Hove level. Elsewhere near the van Hove level, $p$-wave superconductivity is the leading instability, with sizable divergence scale. Notice that $\La_c$ depends on $U$ very sensitively. It changes by more than five orders of magnitude as $U$ is increased from $0.5$ eV to $3.5$ eV (see notations of the color scale). The interaction estimated in first principle calculations is roughly $U\sim 0.7$ eV. This would correspond to a divergence scale $\La_c < 10^{-6}$ eV, questioning the practical interest of the underlying $p$-wave SC. Therefore a more refined estimation of $U$ is needed before one can decide whether the SC state is of practical interest. On the other hand, we have not considered Coulomb repulsion $V$ on the NN bonds so far. A weak $V$ enhances charge-charge interaction at zero momentum. Since this overlaps in part with SDW interaction it could enhance the SDW fluctuations, and SC as a consequence. This effect is shown in Fig.\ref{Tc} for $U=2$ eV. The transition temperature is enhanced significantly from $V=0$ eV (solid line) to $V=0.2$  eV (dashed line), for both SC and SDW orders at the corresponding filling levels. We remark that in the SC phase the pairing amplitude on second-neighbor bonds starts to increase (relative to that on the NN bond) in line with increasing repulsion $V$ on NN bonds. Since we are material oriented we do not consider large values of V, which would drive the system into the CDW state.

\section{Summary} \label{SMR}

We investigated the electron instabilities in the single-sheet BC$_3$ near the van Hove filling. In the weak coupling limit $p$-wave SC is favorable but only below a tiny energy scale. For a moderate Hubbard interaction, the ferromagnetic-like SDW order dominates in the immediate vicinity of the singularity. Elsewhere near the singularity the $p$-wave superconductivity prevails. A small nearest-neighbor Coulomb repulsion can enhance the superconductivity. The wide range of $p$-wave SC regime in the phase diagram is a manifestation of the type-II van Hove singularity. However, the transition temperature becomes practically sizable only if the bare interaction is moderately strong.

\acknowledgments{The project was supported by  National Key Research and Development Program of China (under grant No. 2016YFA0300401) and NSFC (under grant Nos. 11574134 and 11604168).}

\section{Appendix}
 
For self-completeness, here we present necessary technical details for SM-FRG applied in the main text. For brevity we first suppress sublattice and/or orbital labels, to which we will come shortly. Consider the interaction hamiltonian $H_I=(1/2)c_{1\si}^\dagger c_{2\si'}^\dagger \Ga_{1234} c_{3\si'} c_{4\si}$. Here the numerical index labels single-particle quantum numbers, such as momentum/position, and we leave implicit the momentum conservation/translation symmetry. The spin SU(2) symmetry is guaranteed in the above convention for $H_I$. The idea of FRG is to get the one-particle-irreducible interaction vertex $\Ga$ for fermions whose energy/frequency is above a scale $\La$. (Thus $\Ga$ is $\La$-dependent.)
Equivalently, such an effective interaction can be taken as a generalized pseudo-potential for fermions whose energy/frequency is below $\La$. It is useful to  define matrix aliases of the rank-4 `tensor' $\Ga$ via
\eqa \Ga_{1234}=P_{(12)(43)}=C_{(13)(42)}=D_{(14)(32)}.\eea
Here $P$, $C$ and $D$ are matrices of combined indices, reflecting scattering amplitudes for fermion bilinears in the pairing, crossing and direct channels. Starting from the bare interactions at $\La=\infty$, the interaction vertex flows toward decreasing scale $\La$ as,
\eqa \frac{\p \Ga_{1234}}{\p\La} = &&[D\chi^{ph}(D-C)+(D-C)\chi^{ph}D]_{(14)(32)}\nn
&&+ [P\chi^{pp}P]_{(12)(43)} - [C\chi^{ph}C]_{(13)(42)},
\label{Eq:dV} \eea
where matrix convolutions are understood within the square brackets, and
\eqa && \chi^{pp}_{(ab)(cd)} = \frac{1}{2\pi}[G_{ac}(\La)G_{bd}(-\La)+(\La\ra -\La)],\nn
&& \chi^{ph}_{(ab)(cd)} = -\frac{1}{2\pi}[G_{ac}(\La)G_{db}(\La)+(\La\ra -\La)],
\label{Eq:def} \eea
where $G$ is the normal state Green's function, and we used a hard-cutoff in the continuous Matsubara frequency. \\

From $\Ga$ (or its aliases $P$, $C$ and $D$), we extract at a given scale $\La$ the effective interactions in the general SC/SDW/CDW channels
\eqa (V^{\rm SC},V^{\rm SDW},V^{\rm CDW}) = (P, -C, 2D-C). \label{eq:channel}\eea
They are matrices describing scattering of fermion bilinears in the respective channels. Since they all originate from $\Ga$, they are overlapped but are naturally treated on equal footing. The effective interactions can be decomposed into eigenmodes. For example, in the SC channel (with a zero collective momentum),
\eqa
[V^{\rm SC}]_{(\v k,-\v k)(\v k',-\v k')} = \sum_m f_m(\v k)S_m f_m^{*}(\v k'),
\eea
where $S_m$ is the eigenvalue, and $f_m(\v k)$ is the eigenfunction, which can be expanded in terms of lattice harmonics, such as $e^{i\v k\cdot \v r}$ where $\v r$ is the distance between the fermions within a fermion bilinear. We look for the most negative eigenvalue, say $S=\min[S_m]$, with an associated eigenfunction $f(\v k)$. If $S$ diverges at a scale $\La_c$, it signals the instability of the normal state toward a SC state, with a pairing function described by $f(\v k)$. Similar analysis can be performed in the CDW/SDW channels, with the only exception that in general the collective momentum $\v q$ in such channels is nonzero. Since $\v q$ is a good quantum number in the respective channels, one performs the mode decomposition at each $\v q$. There are multiple modes at each $\v q$, but we are interested in the globally leading mode among all $\v q$. In this way one determines both the ordering vector $\v Q$ and the structure of the order parameter by the leading eigenfunction. Finally, the instability channel is determined by comparing the leading eigenvalues in the CDW/SDW/SC channels.\\

For systems with multiple sublattices/orbitals, such as the case of two sublattices in the main text, we take them as general orbits, and the only modification to the above formalism is to take orbital-bilinears into account in the form factors. In this case each form factor is a matrix in the generalized orbital basis.\\

In principle, the above procedure is able to capture the most general candidate order parameters. In practice, however, it is impossible to keep all elements of the `tensor' $\Ga$ for computation. Fortunately, the order parameters are always local or short-ranged. This is notwithstanding the possible long-range correlations between the order parameters. For example, the s-wave pairing in the BCS theory is local, since the gap function is a constant in momentum space. The order parameter in usual Landau theories are assumed to be local. The d-wave pairing is nonlocal but short-ranged. The usual CDW/SDW orders are ordering of site-local charges/spins. The valence-bond order is on-bond but short-ranged. In fact, if the order parameter is very nonlocal, it is not likely to be stable. The idea is, if it is not an instability at the tree level, it has to be induced by the overlapping channel. But if the induced order parameter is very nonlocal, it must be true that the donor channel has already developed long-range fluctuations and is ready to order first. These considerations suggest that most elements of the `tensor' $\Ga$ are irrelevant in the RG sense and can be truncated. \Eq{Eq:dV} suggests how this can be done. For fermions, all 4-point interactions are marginal in the RG sense, and the only way a marginal operator could become relevant is through coherent and repeated scattering in a particular channel, in the form of convolution in \Eq{Eq:dV}. Therefore, it is sufficient to truncate internal spatial range within the fermion bilinear, e.g., between 1 and 2, and between 3 and 4, in  $P_{(12)(43)}$. This means that the form factors are expanded in a truncated set of lattice harmonics. The setback distance between the two groups is however unlimited (thus thermodynamical limit is not spoiled). Similar considerations apply to $C$ and $D$. Eventually the same type of truncations can be applied in the effective interactions $V^{\rm CDW/SDW/SC}$. Such truncations keep the potentially singular contributions in all channels and their overlaps, underlying the key idea of the SM-FRG.~\cite{Wang2012,Xiang2012,Wang2014} The merits of SM-FRG are: 1) It guarantees hermiticity of the truncated interactions; 2) It is asymptotically exact if the truncation range is enlarged; 3) It respects all underlying symmetries, and in particular it respects momentum conservation exactly. 4) In systems with multi-orbital or complex unit cell, it is important to keep the momentum dependence of the Bloch states, both radial and tangential to the Fermi surface. This is guaranteed in SM-FRG since it works with Green's functions in the orbital basis. These are important but may be difficult to implement in the more conventional patch-FRG applied in the literature.~\cite{Honerkamp2001, Metzner2012, Platt2013}\\

To check the convergence of the real-space truncation for fermion bilinears discussed above, we define $L_c$ as the maximal distance between the two fermions within a fermion bilinear. We take a sufficiently large $L_c$ such that the results are not sensitive to a further increase of $L_c$. In the main text, we used $L_c$ up to the third-neighbor bond.

%\bibliography{BC3ref}

\end{document}